# Collective excitations in spin-polarized bilayer graphene

Nguyen Van Men [1,2], Nguyen Quoc Khanh[2,3*], and Dong Thi Kim Phuong[1,2]

[1]University of An Giang - VNUHCM, 18-Ung Van Khiem Street, Long Xuyen, An Giang, Viet Nam (Email: nvmen@agu.edu.vn).

[2]Vietnam National University, Ho Chi Minh City, Viet Nam.

[3]University of Science - VNUHCM, 227-Nguyen Van Cu Street, 5th District, Ho Chi Minh City, Viet Nam (Email: nqkhanh@hcmus.edu.vn).

**Abstract**

We calculate the plasmon frequency $\omega$ and damping rate $\gamma$ of plasma oscillations in a spin-polarized BLG system. Using the long wavelength approximation for dynamical dielectric function, we obtain an analytical expression for plasmon frequency showing that degree of spin polarization $P$ has negligible effect on the long wavelength plasmon frequency. Numerical calculations demonstrate that the plasmon frequency increases (decreases) noticeably (slightly) with the increase in spin polarization in large (small) wave-vector $q$ region. We also find that the damping rate and the shape of $\gamma$ as a function of $q$ depend strongly on $P$. The increase in carrier density decreases significantly both plasmon frequency and damping rate independently of the spin polarization. The numerically calculated critical wave vector, at which the plasmon dispersion curve hits the edge of electron-hole continuum, decreases with $P$ and can be used to experimentally determine the degree of spin polarization.

**1. Introduction**

Collective excitations in materials have lots of technological applications covering fields ranging from energy storage to optical and membrane technology. In the last decades, plasmon in ordinary two-dimensional electron gas (2DEG) systems has been studied and applied intensively. Graphene, a perfect two-dimensional system, is considered as an excellent candidate replacing silicon materials used in recent years because of its unique electrical and optical properties [1-18]. It was shown that low-energy quasi-particles in bilayer graphene (BLG) behave as massive chiral fermions, compared to massive non-chiral (massless chiral) fermions in 2DEG (monolayer graphene (MLG)). Therefore, the density-density response function and dynamical dielectric function of BLG also differ significantly from those of 2DEG and MLG. As a result, screening effects and collective excitations in BLG show also different features [19-26].

In the presence of in-plane external magnetic field, carriers in graphene systems are spin polarized with negligible magneto-orbital coupling due to the negligible thickness of graphene. Previous researches have discovered that the spin polarization can change substantially the characters of 2DEG systems [27-31]. Recently, the authors of Ref. [32] have found remarkable differences in plasmon properties of spin-polarized MLG, compared to those of unpolarized one. It is well-known that plasmons in BLG differ pronouncedly from those in MLG as well as in regular 2DEG due to dissimilarity in the carrier energy dispersion and chirality. Hence, collective excitations in spin-polarized BLG may demonstrate some new interesting features. Up to now, however, no calculations on the collective excitations in spin-polarized BLG have been carried out. Therefore, in this paper, using random-phase-approximation (RPA), we investigate plasmon properties of a spin-polarized BLG system.

**2. Theory**

We consider a zero-temperature, spin-polarized BLG system with a low energy parabolic dispersion relation. The polarization is assumed to be induced by an in-plane Zeeman field $\vec{B}$. Without loss of generality the positive direction of Oz axis can be assumed to be parallel with magnetic field. Under the magnetic field, the energy of electrons in BLG at a given wave vector $\vec{k}$ changes and can be written as [19-20, 32]

$$E_{\vec{k}}^{\lambda,\sigma} = \lambda \frac{\hbar^2 k^2}{2m^*} - \sigma g^* \mu_B |\vec{B}| \tag{1}$$

---

[*] Corresponding author: nqkhanh@hcmus.edu.vn (N.Q. Khanh).

where, $\lambda = \pm 1$ denotes electrons in conduction and valence band, $\mu_B$ is the Bohr magneton, $\sigma = +1(-1)$ indicates spin-up (spin-down) electrons, $g^*$ is the electron Landé g-factor and $m^* = 0.033 m_0$, with $m_0$ being the free electron mass, is the effective mass of electrons in BLG. In order to consider only the upper valence and the lower conduction band below we restrict to the case of carrier density $n \leq 10^{13} cm^{-2}$, corresponding to the Fermi energy $E_F = \hbar^2 \pi n / (2m^*) \leq 0.363 eV$, which is smaller than the minimum of the second conduction band ($\approx 0.4 eV$) [24].

Collective excitations in the system can be obtained from the zeroes of dynamical dielectric function [32-43]

$$\varepsilon(q, \omega - i\gamma) = 0 \qquad (2)$$

In case of weak damping ($\gamma \ll \omega$), the plasmon dispersion and decay rate are determined from the following equations [32-43]

$$\text{Re}\,\varepsilon(q, \omega_p, T) = 0 \qquad (3)$$

and

$$\gamma = \text{Im}\,\varepsilon(q, \omega_p, T) \left( \frac{\partial \text{Re}\,\varepsilon(q, \omega, T)}{\partial \omega} \bigg|_{\omega = \omega_p} \right)^{-1} \qquad (4)$$

Within RPA, the dynamical dielectric function of spin-polarized BLG has the form [32-43]

$$\varepsilon(q, \omega) = 1 - v(q) \Pi(q, \omega) \qquad (5)$$

where $v(q) = 2\pi e^2 / (\kappa q)$, with $\kappa$ being the static dielectric constant of the substrate, denotes the bare Coulomb interaction of electrons in momentum space and $\Pi(q, \omega)$ is the response function of spin-polarized BLG [19-21, 32]

$$\Pi(q, \omega) = \Pi_\uparrow^0(q, \omega) + \Pi_\downarrow^0(q, \omega) \qquad (6)$$

with

$$\Pi_\sigma^0(q, \omega) = \frac{g_v}{L^2} \sum_{\lambda, \lambda', \vec{k}} \left| g_{\vec{k}}^{\lambda, \lambda'}(q) \right|^2 \left[ \frac{f(E_{\vec{k}}^{\lambda, \sigma}) - f(E_{\vec{k}+\vec{q}}^{\lambda', \sigma})}{\omega + E_{\vec{k}}^{\lambda, \sigma} - E_{\vec{k}+\vec{q}}^{\lambda', \sigma} + i\delta} \right]. \qquad (7)$$

In Eq. (7), $g_v = 2$ indicate the valley degeneracy, $f(x)$ is the Fermi – Dirac function, and

$$\left| g_{\vec{k}}^{\lambda, \lambda'}(q) \right|^2 = \frac{1}{2} \left[ 1 + \lambda \lambda' \cos(2\theta_{\vec{k}} - 2\theta_{\vec{k}+\vec{q}}) \right] \qquad (8)$$

is the overlap function.

In long wavelength limit ($q \to 0$), the spin-resolved response function of spin-polarized BLG has simple form [19-21]

$$\Pi_\sigma^0(q, \omega) = \frac{g_v m^*}{\pi} \left( \frac{\hbar^2 k_F}{2m^*} \frac{q}{\omega} \right)^2 (1 + \sigma P) \qquad (9)$$

Here $k_F = \sqrt{\pi n}$ is the Fermi wave-vector of unpolarized BLG with carrier density $n = n_\uparrow + n_\downarrow$, and $P = (n_\uparrow - n_\downarrow)/(n_\uparrow + n_\downarrow)$, with $n_\sigma$ being the spin-polarized electron density, denotes the spin-polarization. In this approximation, the solution of Eq. (3) can be easily obtained,

$$\omega^2 = \frac{2e^2 g_v E_F}{\kappa} q \qquad (10)$$

where $E_F = \hbar^2 k_F^2 / (2m^*)$ is the Fermi energy of unpolarized BLG.

Eq. (10) shows that the plasmon frequency in spin-polarized BLG is almost independent of spin polarization in long wavelength limit. It is well-known that the plasmon mode is undamped until the dispersion curve enters the single-particle-excitation (SPE) continuum, determined by equations $(\omega_1/E_F) = (q/k_F)^2 + 2(q/k_F)\sqrt{1+\sigma P}$ and $(\omega_2/E_F) = (q/k_F)^2 - 2(q/k_F)\sqrt{1+\sigma P} + 2(1+\sigma P)$, at a critical wave-vector $q_c$. In case of spin-polarized BLG $q_c$ is determined by the intersection of plasmon dispersion curve with the lower edge of the inter-band continuum of the minority spin carriers. In the limit $q \to 0$, using Eq. (10) we have

$$q_c = k_F \left[ p + \sqrt{-\frac{p^2}{3} + \frac{2(p^4 - 3r_s p)}{3\Delta(r_s, p)} + \frac{\Delta(r_s, p)}{6}} - \frac{1}{2}\sqrt{-\frac{8p^2}{3} - \frac{8(p^4 - 3r_s p)}{3\Delta(r_s, p)} - \frac{2\Delta(r_s, p)}{3} + \frac{4\sqrt{6}r_s}{\sqrt{-2p^2 + \frac{4(p^4 - 3r_s p)}{\Delta(r_s, p)} + \Delta(r_s, p)}}} \right] \qquad (11)$$

where $p = \sqrt{1-P}$, $r_s = e^2 g_v m^* / (\kappa k_F)$ and

$$\Delta(r_s, p) = \left[ -8p^6 + 36 p^3 r_s + 3\left(9 r_s^2 + \sqrt{-96 p^6 r_s^2 + 408 p^3 r_s^3 + 81 r_s^4}\right) \right]^{1/3}.$$

In limit of $P \to 0$ (unpolarized BLG), Eq.(11) reduce to

$$q_c|_{P\to 0} = k_F \left[ 1 + \sqrt{-\frac{1}{3} + \frac{2(1-3r_s)}{3\Delta(r_s,1)} + \frac{\Delta(r_s,1)}{6}} - \frac{1}{2}\sqrt{-\frac{8}{3} - \frac{8(1-3r_s)}{3\Delta(r_s,1)} - \frac{2\Delta(r_s,1)}{3} + \frac{4\sqrt{6}r_s}{\sqrt{-2 + \frac{4(1-3r_s)}{\Delta(r_s,1)} + \Delta(r_s,1)}}} \right]$$

(12)

where $\Delta(r_s, 1) = \left[ -8 + 36 r_s + 3\left(9 r_s^2 + \sqrt{-96 r_s^2 + 408 r_s^3 + 81 r_s^4}\right) \right]^{1/3}$.

It is seen from Eq. (12) that the dependence of critical wave-vector $q_c$ on interaction parameter $r_s$ is quite complicated, compared to that in the MLG case given in Ref. 32. Note that in case of MLG $r_s$ is independent of carrier density $n$ hence $q_c$ depends only on dielectric constant $\kappa$. In case of BLG howevere $r_s = e^2 g_v m^* / (\kappa k_F)$ $\sim n^{-1/2}$ implying that $q_c$ is a function of carrier density as shown belove in Fig. 6(b).

## 3. Results and discussions

In this section, we present our numerical results for plasmon frequency and damping rate of spin-polarized BLG system with spin polarization $P$ and total doped electron density $n$. We have used Eqs. (3)-(17) of Ref. 20 with $g$ = 2 and dimensionless variables $x = k/k_{F_\sigma}$, $y = q/k_{F_\sigma}$ and $z = \omega/E_{F_\sigma}$ to calculate numerically the response function $\Pi_\sigma^0(q, \omega)$ of spin polarized BLG.

Figure 1 illustrates plasmon frequency (a) and damping rate (b) in a BLG system with $n = 10^{12} cm^{-2}$ and $P = 0.5$ for two different substrates $SiO_2$ ($\kappa_{SiO_2} = 3.8$) [25] and $BN$ (boron nitride, $\kappa_{BN} = 5.0$) [44]. We observe that plasmon mode is undamped until the dispersion curve hits the edge of the single-particle-excitation (SPE) of the minority spin carriers (thin dashed-dotted lines) at $q \approx 0.1 k_F$. In the SPE interband continuum of minority spin carriers the damping rate increases, reaches a peak and then decreases before the dispersion curve enters the SPE inter-band continuum of majority spin (thick dashed-dotted lines) carriers at $q \approx 0.54 k_F$. At larger wave-vectors damping rate increases again before decreasing to zero as plasmon approaches the intra-band continuum and vanishes. The increase in dielectric constant of substrate decreases significantly the plasmon frequency and damping rate.

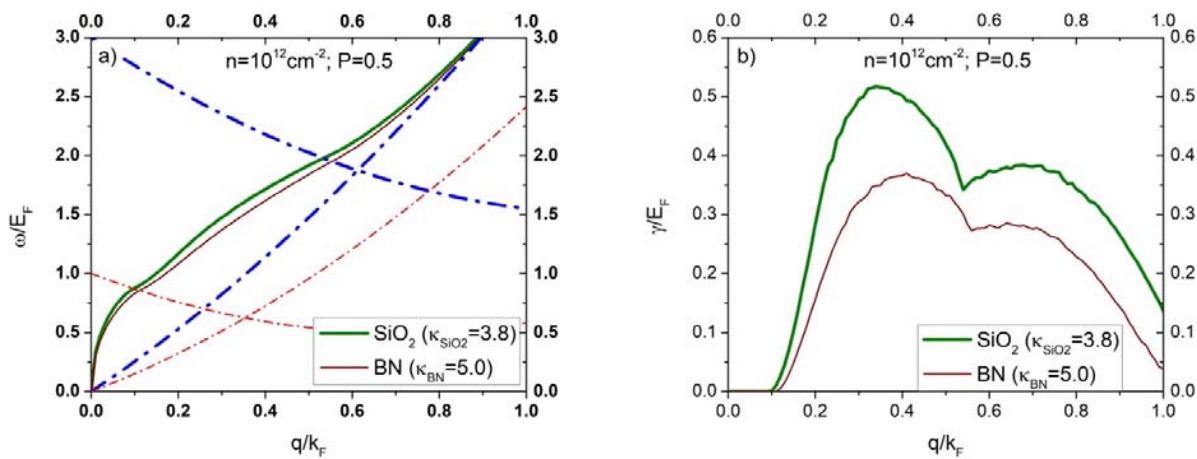

**Fig. 1.** Plasmon frequency (a) and damping rate (b) in spin-polarized BLG with $n = 10^{12} cm^{-2}$, $P = 0.5$ for two different substrates $SiO_2$ ($\kappa_{SiO_2} = 3.8$) and $BN$ ($\kappa_{BN} = 5.0$). Thick (thin) dashed-dotted lines show the boundaries of SPE continuum of majority (minority) carriers.

In order to see the effects of spin polarization on plasmon modes in spin-polarized BLG, we plot in Fig. 2 plasmon frequency (a) and damping rate (b) as a function of wave-vector for $n = 10^{12} cm^{-2}$ in two cases $P = 0$ and $P = 0.5$. The dotted line corresponds to analytical results given in Eq. (10) with the same parameters. As seen from Fig. 2(a), the analytical results have a good agreement with the numerical ones in long wavelength region. For sufficiently small wave-vectors, plasmon curves in both cases $P = 0$ and $P = 0.5$ are identical. Numerical results also indicate that in long wavelength limit plasma frequency shows a weak dependence on the spin polarization. We observe that when the wave vector increases slightly, the plasmon frequency in spin-polarized BLG is smaller than that in unpolarized case. However, at larger wave-vectors, the spin-polarization $P$ increases significantly plasmon frequency compared to the case $P = 0$. The plasmon curve of spin-polarized BLG merges the SPE boundary similarly as in unpolarized systems [20]. Fig. 2(b) shows that the plasmons became damped at much smaller $q$ due to spin-polarization and plasmon damping rate of unpolarized system shows no kink as already known.

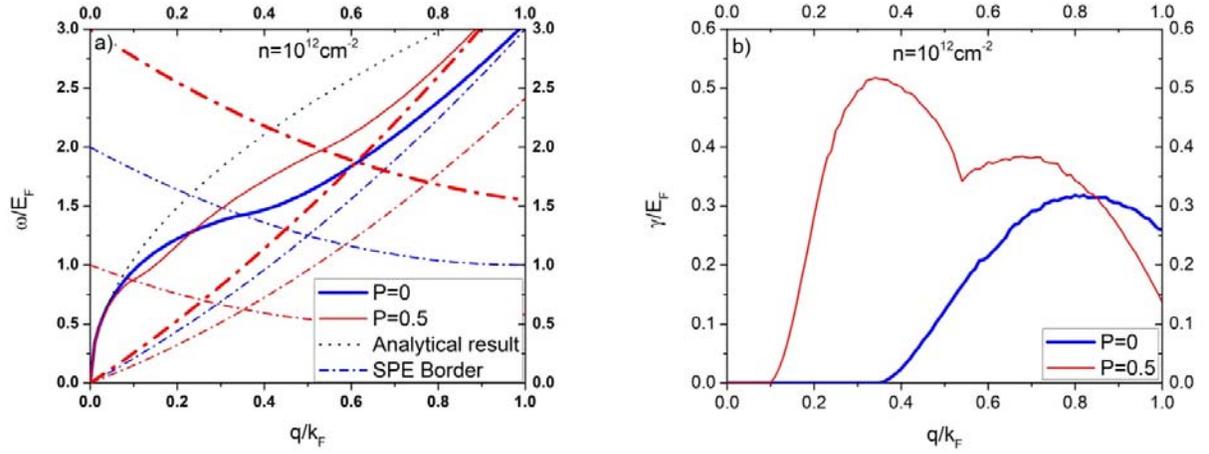

**Fig. 2.** Plasmon frequency (a) and damping rate (b) in spin-polarized BLG ($P = 0.5$) and in unpolarized BLG ($P = 0$) for $n = 10^{12} cm^{-2}$. Thick (thin) dashed-dotted lines show the boundaries of SPE continuum of majority (minority) carriers.

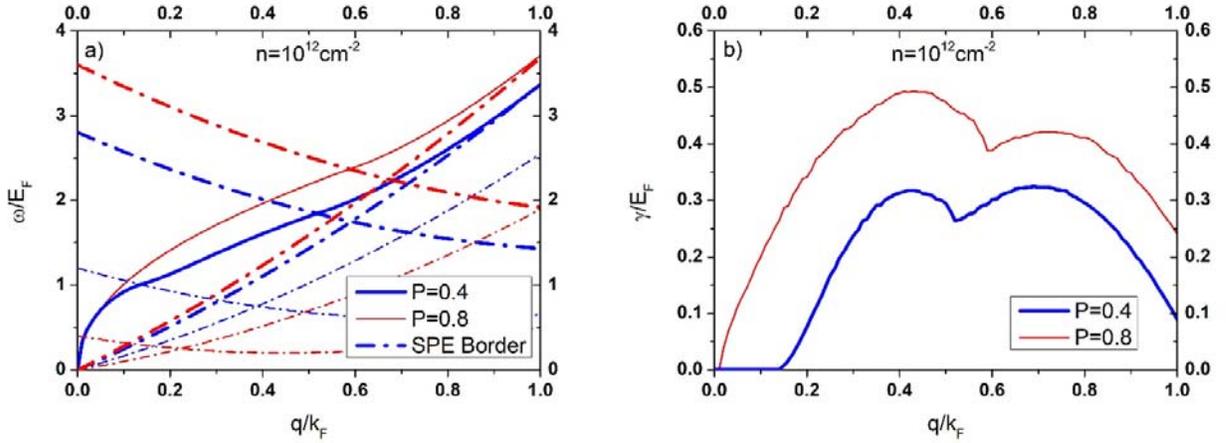

**Fig. 3.** Plasmon frequency (a) and damping rate (b) for spin-polarized BLG with $n = 10^{12} cm^{-2}$ for $P = 0.4$ and $P = 0.8$. Thick (thin) dashed-dotted lines show the boundaries of SPE continuum of majority (minority) carriers.

In Fig. 3 we compare the plasmon frequency (left) and the damping rate (right) of plasma oscillations in partially polarized BLG with carrier density $n = 10^{12} cm^{-2}$ for two cases $P = 0.4$ and $P = 0.8$. As seen from the figure, for not too small wave-vector $q$, an increase in spin-polarization leads to a remarkable increase in plasmon frequency. In addition, Fig. 3(b) indicates that the energy loss in case $P = 0.8$ occurs at much smaller wave-vector, about $q \approx 0.01 k_F$, compared to $0.15 k_F$ in case $P = 0.4$ and the damping rate in the case of $P = 0.8$ is larger and has a local minimum at larger $q$ in comparison with the case $P = 0.4$. It is easily understood because as $P$ increases the number of minority (majority) carriers decreases (increases), the lower edge of the inter-band continuum of the minority (majority) spin carriers moves down (up) and the unoccupied states of minority carriers are larger, leading to stronger interband transitions. In both two cases of polarization, as wave-vector increases, plasmon curves approach the boundary of intra-band SPE continuum and disappear while the damping rate decreases to zero.

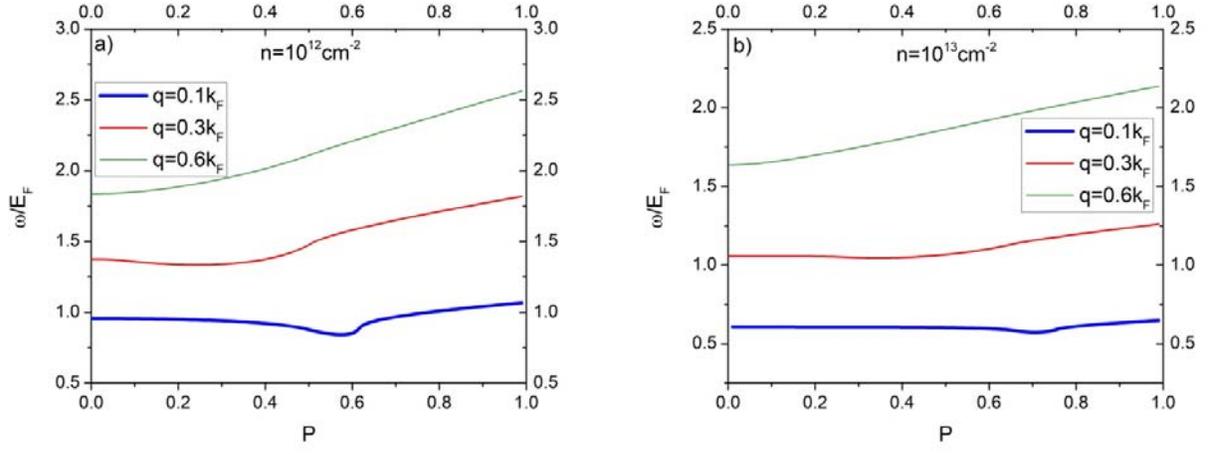

**Fig. 4.** Plasmon frequency as a functions of polarization $P$ in spin-polarized BLG with $n = 10^{12} cm^{-2}$ (a) and $n = 10^{13} cm^{-2}$ (b) for several wave-vectors.

For more information about the effects of spin polarization on plasmon modes, we plot in Fig. 4 the plasmon frequency as a function of spin polarization for several values of wave-vector in two cases $n = 10^{12} cm^{-2}$ and $n = 10^{13} cm^{-2}$. It is seen from the figures that the increase in degree of spin polarization $P$ increases slightly (strongly) plasmon frequency at small (large) wave-vectors and plasmon frequency at large wave-vectors fluctuates with $P$ much more than that at small $q$.

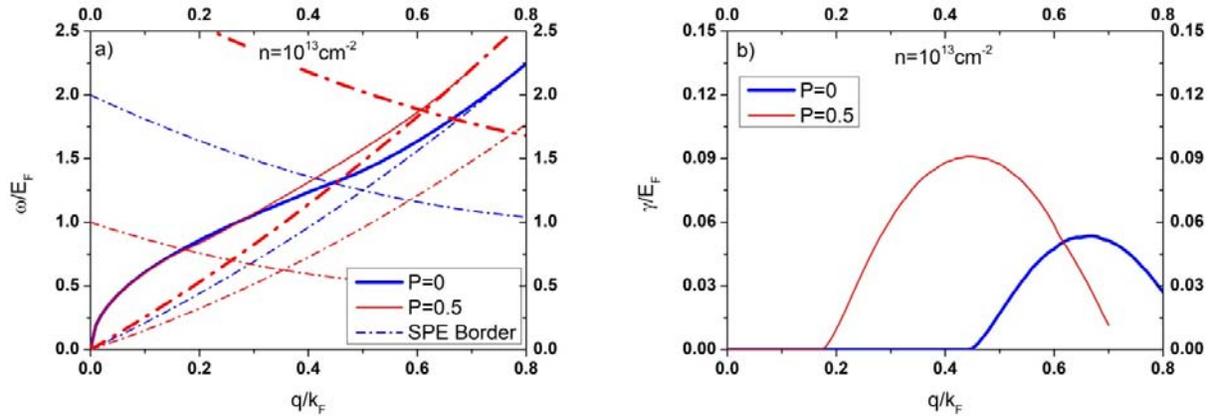

**Fig. 5.** Plasmon frequency (a) and damping rate (b) in spin-polarized BLG with $n = 10^{13} cm^{-2}$ for $P = 0.5$ and $P = 0$. Thick (thin) dashed-dotted lines show the boundaries of SPE continuum of majority (minority) carriers.

We now consider the effects of carrier density on plasmon characters in BLG. Fig. 5 demonstrates plasmon frequency (a) and the damping rate (b) in spin-polarized and unpolarized BLG with $n = 10^{13} cm^{-2}$. Figs. 2(a) and 5(a) indicate that in both two cases $P = 0$ and $P = 0.5$ the increase in electron density decreases pronouncedly plasmon frequency. We also find from Fig. 5(b) that at large carrier densities the plasmon dispersion curve merges the upper edge of the intra-band continuum of majority spin carriers and vanishes before entering the SPE region of minority component. Hence for $n = 10^{13} cm^{-2}$ the damping rate shows no kink as a function of $q$ as in the case $n = 10^{12} cm^{-2}$.

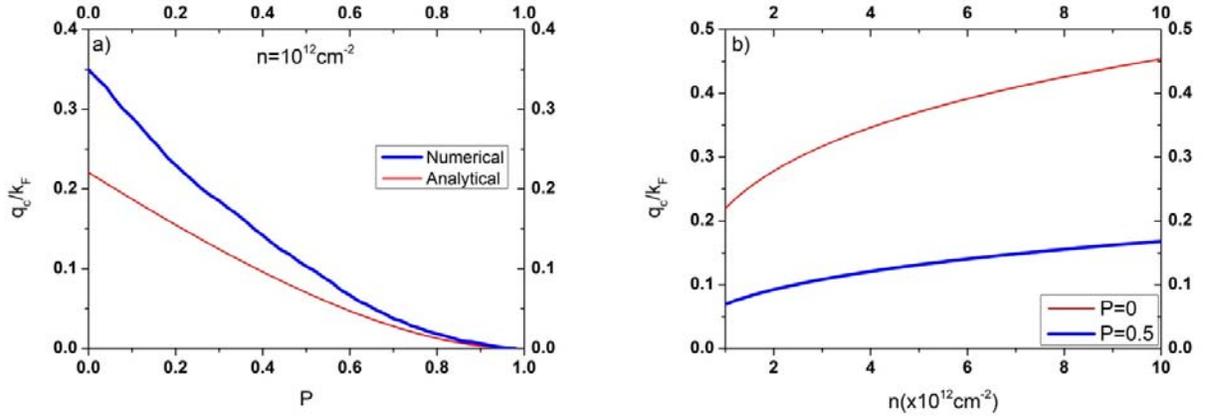

**Fig. 6.** The critical wave-vector $q_c$ at which the plasmon dispersion curve hits the edge of the SPE of the minority carriers as a function of polarization (a) and carrier density (b).

In Fig. 6 we show the analytical and numerical critical wave-vector $q_c$ as a function of degree of spin polarization (a) and carrier density (b). As seen from Fig. 6(a), the critical wave-vector decreases with increasing spin polarization because the lower edge of the inter-band continuum of the minority spin carriers ships down as $P$ increases. For smaller $P$ the lower edge of the inter-band continuum of the minority spin carriers is higher and $q_c$ is larger. Hence the long wavelength limit result for $q_c$ is less correct and differs remarkably from the numerically calculated critical wave-vector. Note that our numerical value of $q_c$ can be used to estimate the degree of spin polarization experimentally [32]. Fig. 6(b) demonstrates that the increase in carrier density $n$ increases the critical wave-vector in both two cases $P=0$ and $P=0.5$. In addition, in the unpolarized case the critical wave-vector depends on $n$ more strongly than in the polarized one.

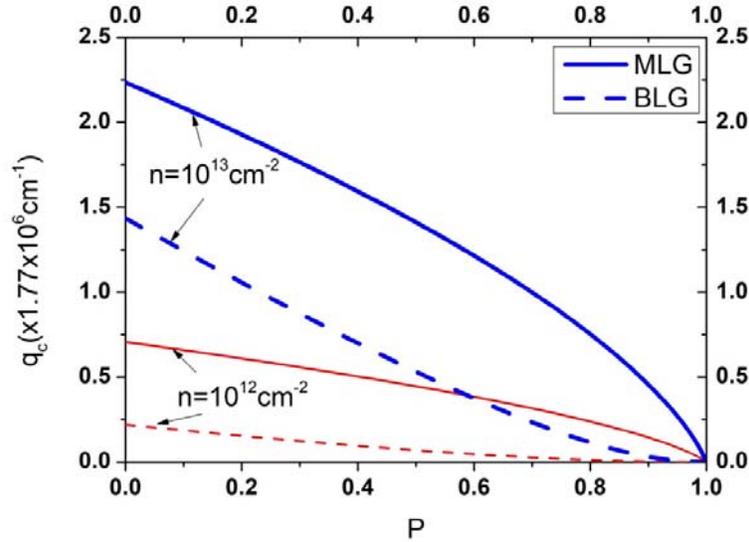

**Fig. 7.** The critical wave-vector $q_c$ at which the plasmon dispersion curve of MLG and BLG hits the edge of the SPE of minority carriers for $n=10^{12}\,cm^{-2}$ and $n=10^{13}\,cm^{-2}$.

Finally, we note that the differences in collective excitations of unpolarized BLG and MLG systems have been shown by several authors [20, 36, 37]. In order to make a comparison between plasmon properties of spin-polarized BLG and MLG we plot in Fig. 7 the critical wave-vector $q_c$ as a function of spin polarization in two cases $n=10^{12}\,cm^{-2}$ and $n=10^{13}\,cm^{-2}$ for both graphene systems. The figure indicates that $q_c$ behaves as a decreasing function of polarization and decreases more quickly for higher $n$. For both MLG and BLG, the increase in carrier density increases remarkably the critical wave-vector at which the plasmon dispersion hits the boundary of the SPE of minority carriers. In addition, in spin-polarized systems, at a given carrier density the critical wave-vector for MLG is much larger than that for BLG with the same parameters

**4. Conclusion**

In summary, the plasmon frequency and damping rate of plasma oscillations in a spin-polarized BLG system at zero-temperature have been calculated for the first time. Obtained analytical and numerical results indicate that plasmon frequency is dispersing as $\sqrt{q}$ and almost independent of spin polarization in long wavelength limit. The plasmon frequency and damping rate decreases significantly with increasing substrate dielectric constant. We also find that the increase in degree of spin polarization increases slightly (strongly) plasmon frequency at small (large) wave-vectors. The maximum value of damping rate is larger for smaller $P$ due to larger number of unoccupied states of minority carriers, leading to easier inter-suband e-h excitations. For large carrier density ($n \sim 10^{13} cm^{-2}$) the damping rate of partially polarized system shows no kink as a function of $q$ similarly as in unpolarized case. However, in low density systems ($n \sim 10^{12} cm^{-2}$), $\gamma$ of partially polarized BLG shows a local minimum when the plasmon curve hits the lower edge of the interband continuum of majority spin carriers. We have also calculated analytically and numerically the critical wave-vector $q_c$ which can be used to determine the degree of spin polarization experimentally.

**Acknowledgement**

This research is funded by Vietnam National Foundation for Science and Technology Development (NAFOSTED) under grant number 103.01-2020.11.